\documentclass{article}
\usepackage{amsmath,amsfonts,amssymb,mathrsfs,graphicx}
\begin{document}
\title{Agent Components and the Emergence of Altruism in Social Interaction Networks}

\author{Fariel Shafee \\
Department of Physics\\ Princeton University\\
Princeton, NJ 08540\\ USA.}
\date{23 January 2009}
\maketitle

\begin{abstract}
We discuss a special aspect of agents placed in a social network.  If an agent can be seen as comprising many components, the expressions and interactions among these components may be crucial.  We discuss the role of patterns within the environment as a mode of expression of these components.  The stability and identity of an agent is derived as a function of component and component-pattern identity.  The agent is then placed in a specific social network within the environment, and the enigmatic case of altruism is explained in terms of interacting component identities.

\end{abstract}

\section{Introduction}
The field of complexity has recently gained a lot of attention because of its ability to explain many complicated problems in terms of the organization of simpler units.  Hierarchical structures existing to form levels of organizations were proposed (see Ahl and Allen, 1996).  Interactions among units within a complicated organization was also modeled in terms of biological processes, such as the function of neural networks (see Hopfield and Herz 1995), the behavior of the stock market (Black and Scholes 1973), traffic flows (e.g. Helbing and Schreckenberg 1999) etc.  Social organizations have also been studied to some extent (e.g. Newman et al 2006, Reichardt and White 2007).  In our previous work (Shafee 2008), we have proposed a model based on agent components and interactions among agent-components, and identities of agents and social groups deriving from such interactions as minima in the interaction potential landscape.  In another paper, the role of perceptions and beliefs, and interesting dynamics deriving from the connections between such beliefs within a network were discussed.

In this work, we extend some concepts associated with the model in terms of components and the perception and expression of components.  We then present an example of a specific case when components of an agent and components of some coherently oriented agents interact to preserve a local potential minimum in order to explain the puzzling social phenomenon of altruism.

\section{Interaction-based Social Model: Brief Review}

Previously (Shafee 2004, Shafee 2008) an agent was modeled as an array of states, that were expressed by means of weights assigned to them.  The satisfaction or utility derived was expressed as the value of the potential of interactions between the agent's states ad states found in other agents and the environment.

The model was inspired by spin glasses (Edward and Anderson 1975 , Sherrington and Kirkpatrick 1975) where interactions among neighboring spins contributed to the total interaction potential of the system.  The system itself evolved towards minimum potential attractors.

An agent's interactions were split into three major categories to form three components of the interaction potential:

\begin{equation}
H_{potential} = H_{self}+H_{agent-agent}+H_{env}
\end{equation}

Here the respective Hamiltonian components have interaction terms of the form
\begin{eqnarray}
H_{self}= - J^{ab}_i s_i^a s_i^b\\
H_{env} = -J^{a}_ih s_i^a . h^a \\
H_{agent-agent} = J^{ab}_{ij} s^i_a s^j_b
\end{eqnarray}

The $J$'s are the coupling constants, representing the strength of an interaction. $i$ is the local agent, and $j$ is a neighboring agent. $a$ is the label assigned for  a characteristic variable.

Agent variable states are denoted with $s$'s and in the spirit of the original spin glass model, larger environment effects are expressed with $h$'s as environment field components.

Each agent interacts to minimize his own interaction potential to maintain his aggregate identity.

In the previous paper (Shafee 2008), the basic concepts of the agent-interaction model was discussed and social structures and dynamics derived from inter-agent interactions were studied. Here the identity and weight of agent-components are analyzed un detail, and an extreme case of altruism arising from component identity within a social, and agent identity is modeled.

\subsection{Matching of Components and Symbiosis}
Interaction terms may show co-operative couplings among diverse components within the agent and also among matching components of different agents. The symbiotic couplings allow the states to {\it seek one another} in a highly correlated manner, while matching components among agents find favorable interactions just like spins aligned in the same direction are energetically favored in magnetism.

\section{Components, Patterns and Propagation}

In our previous work (Shafee 2008), we have argued how inter-agent interactions giving rise to fuzzy emotions and identities at various levels are based on sums of interacting components.  Here, we discuss how these components depend on coupling matrices, giving rise to certain patterns.
It is known that for complex systems, the organization of the system comprising similar units yield interesting behaviors (Hopfield and Herz 1995).  For example, in the complex human brain almost identical neurons are connected together.  All these neurons behave according to more or less the same rules.  However, their topological organization result in different personalities (Peterson and Carson 2000) that are genetically fixed, although a subset of neurons can update their physical correlation in response to agent-environment interactions, producing memories local to the agent (see Lebedev et al 2005  for a study of neural adaptivity).  Hence, the agent-environment interface is included in the agent's extended identity.  The preferences of agents derive from the firing patterns of the assembly of neurons, which is independent of the exact neuron producing the pattern.  If these preferences are taken as variable states of an agent, it is the expression of these firing patterns that interact with the environment and not the exact neuron.  Again, if the gene determines the topological organizations of certain parts of the brain associated with certain behaviors or preferences, another agent with the same gene pattern and not the same gene would produce similar behavioral patterns that the agent can identify with. In the material universe although the proportion of elementary particles (e.g. the proportion of neutrons and protons) in a molecule making a gene segment is fixed, reorganizing the same number of protons and neutrons would in general produce different patterns at the macroscopic level.

At the agent-environment interface, skin cells are constantly lost.  Similar but not exact replica cells are created to replace the lost ones, though the rate of regeneration of skin cells programmed into the internal mechanism of the body may not be equal to the exact number of cells lost in a one-to-one basis.

\subsection{Stable and Average-Stable Components}
Neurons related to memory cannot regenerate.  The exact organizations of neurons in the form of weights connecting the neurons give rise to the memories.  The memories themselves depend on the agent-environment interactions, which is not determined by the genetic pattern, and hence loss and regeneration of neurons would erase any memory accumulated.  The extensive number of interconnected neurons are thus preserved within a delicate environment protected within organs which themselves can regenerate continuously against certain degrees of losses, so that in an aggregate they are able to maintain an ``environment field" necessary to preserve the nerve cells. Hence, as in our previous work (Shafee 2008) agents were placed with respect to environment fields, neurons can as well be said to be in connection with optimal fields produced by structures in their environment.  The complexity of an agent derives from being able to protect multitudes of data and states within the many precise variables of exact organizations of  neural units protected by an average constant field that is produced by an average aggregate of cells.  So, it is not the total number of cells or the total number of types of cells organized into separate organs that produce the complete measure of complexity.

When a perception organ is considered, or a limb that directly interacts with the environment, the current state of the organ is dependent on the external field and the internal signals.  The reorganization or disturbance of the organ allows the agent to receive information, or to send information to be stored.  These organs can also reorganize into specific states such that interactions with such organs can force the environmental states to reorganize or dissociate into smaller components by means of transfer of energy. When an organ as a whole is interacting with an environment, the individual subunits are also interacting with smaller subunits of the environment.  For example, friction would cause smaller units of the organ to be lost.

The organization of the organ, which is a function of each subcomponent's location and inter-relation, would depend on the internal and external environment the organ is connected with, since the resultant equilibrium of any unit derives from the balancing forces acting on it, with the magnitude, direction and interaction range of each force taken into account.  The stochastic nature of many degrees of freedom associated with the environment makes it most probable that the subcomponents of the local environment would reorganize, slightly changing the balance of force at a certain point as fluctuation, and a replenished part may only have approximately the same organization.

\subsection{Expression Stability and Weights}
The relationship between a changing memory and, hence, beliefs derived from experience, and the interaction of a genetically defined topology contributing to certain personality traits (Toga and Thompson 2005) preserved together locally as possible output behavioral patterns was recently discussed (Shafee 2008). While the intricate connections among the brain cells and the interrelatedness of the individual cell components make it impossible to make brain transplants, many other components may be substituted within a range of fuzzy matching of components in connections with the couplings with respect to the other ``individualized" organs and components.  For example, blood can be transfused if the specific blood group is identified, and organs like kidneys and hearts can also be transplanted given further detailed matches.  The match of livers has a high probability among close genetic relatives (see Finn 2000 for a discussion of organ matching and donation). A mismatched organ causes rejection and often death. Although the notion of cellular memory remains controversial, hormones have been found to be intricately related with personalities and preferences (see  e.g. Alder 1986). Injecting hormones or transplanting or severing an organ that produces certain hormones can change the personality.

As was briefly mentioned in  (Shafee 2008), the human genome has been found to consist of genes that are 99.9 percent identical within the entire genetic pool (Walton 2004). However, not all segments of the gene have the same degree of expression, and some might as well be suppressed.  Only a small fraction of the human genetic material is known to code for protein synthesis (see ??????????).

Small segments of genes can also be related with high degrees of visible or expressed attribute differences. Again, a gene has been shown to express certain exons at certain times leading to the same gene coding for a large family of proteins by means of alternative splicing (Brett et al 2001). Cellular and environmental conditions such as stress, pH etc have also been seen to act as triggers in alternative splicing (Stamma et al 2005).
 It has also been seen that the promoter for a certain gene need not be located on the same gene, hence making it necessary for another gene to exist to trigger the action of one gene (Stamma et al 2005).

Organs specialize in carrying out the instructions of specific chunks from the gene code depending on their diversification.  The development into specialized cells is again brought about by the immediate environment of the cell during growth, which might as well reflect the behavioral pattern of the mother during natation that are not coded within the gene of the child, and the location of growth (see e.g. Nicolopoulou-Stamati 2007).  Different types of muscle cells have also been known to produce different types of acto-mycin because of alternative splicing triggering different expressions of the same gene in different organs (Stamma et al 2005).  The concentration of certain chemicals present in within the body because of the action of genes expressed in one organ can also induce alternate splicing mechanism of another gene in another organ, hence causing one organ to effect another (Stamma et al 2005).

The expression of the internally coupled organs, and at lower hierarchical level cells that make up the organs, and even at further lower level, the genes that are expressed in the behavior of each cell that are coupled, in turn depend on the consistency, degree of coupling and coordination, and hence the fitting of components within a semi-stable compatible structure.  The degree of expression of a certain gene component is dependent on the degree of coupling of that component with other components by means of the couplings of the cells that express the genes, and the couplings of the organs that make the body, and while certain couplings (e.g. those of nerve cells) are expressed intricately causing large shifts of patterns due to small changes, some only produce an aggregate behavior and may have a larger degree of robustness because of internal corrective couplings, so that any error in one is offset by another component adjusting accordingly.

Hence, the mapping of the gene, which is the blueprint of the agent, is not linearly expressed in the agent, and the expression of the agent, based on its local state is a function of both the gene and the environment, with various weights.  The weight of a component (Shafee 2008) is, therefore, related to the expression of the component, which may be a function of the coupling of the component to the number of subcomponents at each level, and the coupling of the coupled subcomponent with the environment's components as well.  The match of two weighted components, and hence the compatibility is expressed is given by the coupling constant, which produces an interaction energy potential. And the time averaged weight of a certain variable is related to the total expression, which takes into account the frequency of the triggering of the action as well. So, the weight of a component on an agent also reflects the ability of minuscule update of the component state to change the total interaction potential within the agent, and is a measure of the priority of the component.

\subsection{Propagation of Patterns and Long Term Perpetuation}
The coupled components and, hence, variables within the agent, with various degrees of stability, matching and interconnected symbiosis, thus produce a semi-stable local interaction potential minimum with respect to the environment.  The idea of the semi-closed nature of such {\it identities} was discussed in (Shafee 2007).
The degree of stability of such identities derive from its being able to regenerate lost components in a complex manner, and its ability to interact to reduce damage.  Hence, the perpetuation of a complex agent identity is dependent on its ability to express variables to diminish the effects of unfavorable changes of the interaction potential (raising of the local minimum) by triggering expressions that ``block" or flip (reorganize or align to a favorable direction) the damaging variable.
This propensity of maintaining a local interaction minimum, by means of strongly coupled components that can be expressed as agent states, can thus be succinctly related to the agent's tendency to perpetuate itself, and hence the patterns that lend the agent its identity.

We consider two possibilities for expression of the genetic code.  One is by means of the expression of a specific small or large chunks of genetic sequences separately, and the other is by the organization of codes so that only certain codes are triggered and expressed as intricately connected sequences, while others may remain dormant depending on the organization.

During meiosis, crossover between chromosomes make chunks of interconnected gene alleles, that can produce degrees of genetically formed skill aptitudes recombine (see  Griffith et al (1993) for a discussion of linked gene mapping based on recombination).  Hence, specific talents and traits that are results of intricate connections of large number of subsegment patterns are often lost.  In a large population, these fragmented sections may recombine, subject to rules of probability, or similar patterns may arise because of the repetition of some genetic code sequences in different segment.  If a characteristic is specific to an average of a sequence, approximate characteristics can be re-expressed even if the exact same segment of sequences is not recombined.

 From the point of view of perpetuation of patterns, some of these ``lost" expressions can to some order approximately reappear in a large gene pool making random genetically unrelated agents share common attributes with certain frequencies.

However, small mutations can cause specific changes in the expression of a characteristic at a large scale due to the connectedness of the specific attribute with other attributes.
Such small mutations are restricted to genetic kinship (given the very small probability of the exact same point mutation happening in a separate case, because the vast number of possible locations for such a mutation within the gene). The loss and perpetuation of such characteristics are subject to the propagation frequency of the agent and the number of genetic kin emerging from the mutated agent.  Hence, though the perpetuation of many components of an agent is related to maintaining a random large gene pool, the perpetuation of these point mutations is subject to maintaining a subcluster of genetically related kin.

Again, though these point mutations can continue over a very long period and many generations, some other generational fixed attributes derive from the lack of reorganization of the Y chromosome and the mother's mitochondrial DNA.  If any arrangement of sequences in these chromosomes produces an expression pattern, they can thus be seen to be perpetuated from one generation to another.  The weights of these specific genetically fixed expressions as opposed to the weights of the lost expressions among kin but available in random unrelated agents would come as a cohesive force of genetic kinship based clustering.

 A faith, or a rule derived from perception, though stored as small weights within a set of  specific neurons, can express itself as a set of highly weighted behavioral patterns.  If the expression of such a faith is related to more historical experiences, the faith gains mass (``inertia") because of its inter-connections (see Shafee 2008b).  Clauses of these faiths may be coupled with genetic attributes and also personal historical experiences. The fixed nature of these coupled terms can also make the faith massive and/or weighted by induction (see Shafee 2008 for such a mechanism). Some faith clauses may be genetically independent, but related to a common highly weighted massive term or a threat (see Shafee 2008b).

The faith itself may have large behavioral consequences, and programming the same faith within a large number of agents by means of common experiences and communication from credible sources (see Shafee 2008b for some possible modes of spread of axioms or faith). A highly weighted faith may exist in different conflicting states among isolated clusters because of the local origin and spread of the faith. The history experienced by the clusters separately that are coupled to their specific faiths, and are thus massive (having a significant inertia or reluctance to change) within the clusters lend the faith its own identity spreading within each cluster.  Hence, within the cluster level, faith-based identities exist, and conflicting states may exist in two separate clusters.

\section{Pattern Recognition}
In neural networks, patterns can be recognized by programming attractors. Yet, the neural networks are given a capacity so that the number of patterns that can be stored is limited.  The attractors allow for the input patterns to be driven to one of these attractors even if the input and the stored patterns are not identical. In a similar spirit, we can define the ``matching" of states at each level, so that the agent recognizes a match or a conflict of another agent's attribute at each level of expression.  The scales related to each of these levels dictate which level is expressed in an interaction. Hence, subtle differences between two states due to different lower level components existing within them may be {\it blurred} out at a different scale, where the recognition of pattern depends on stored matches or mismatches.

\subsection{Pattern Component Identity within the Environment}
As was explained above, while the agent attains his identity by means of the local potential minimum formed by means of interactions between coupled states, the expression of the states of the agents within the environment are by means of re-alignments or organizational changes within the environment.  These changed alignments represent the relative optimality between the agent and the environment, and the stiffness to change or inertia (Shafee 2008) of the agent and the environment. The connection with the environment slowly brings about changes in an agent involving change in values, adaptation and survival of the fittest. The last mentioned  depends on being able to maintain the fit components coherently, given the stochasticity and fluctuation produced by the large degrees of freedom of the environment and the diverse organization of the components. The need for symbiosis among the organs to produce coherent sustainable couplings, and hence the need to squeeze diverse states within a small space is offset by the restriction in the number of variables that can be expressed (Shafee 2008).  The large scale of the environment systems to produce ``fields" optimal for each agent component is offset by the restriction on the possible number of such different fields, each optimal for a different agent component to exist together within the same local environment. The connection of the local environment to a global environment, where other agents with slightly different optimality conditions are connected, causes the local fields to change, and the identity of an agent is by means of maintaining a local environment that sustains his own coupled components to create a local potential minimum.  Hence, the cost of maintaining a local environment optimal to an agent's existence is in producing a change in the connected global environment if the agent were not present. The degree of change in global couplings in order to maintain a local stable pattern of states and fields is  a measure of the expression of the identity of the agent within the context of the environment.

 A one time change in the local environment to produce a state optimal for an agent component is interacted with by other environment systems and agent states coupled to that.  So, the stability of a system depends on the degree of connections of states and components that the particular state is interacting with, and the number and coupling strengths of the states that tend to realign the optimal state in an unfavorable direction. If several agents have the same optimal environments, they will tend to orient their local environment along the direction of the same preference, and the overlapping components can be shared.  The non-overlapping components can be extended to form a larger state where the same alignment exists along that preference, and the complementary  {\it skills} of the multiple agents sharing the same preference can be used to realign the various types of unfavorable interacting environment states.  The connection of multiple agents with the same preference enables multiple states of the agents to be expressed within that same alignment region, decreasing the constraint of the number of expressed states available to shield an agent.  This large scale alignment of the environment along a particular preference thus allows multiple agents to collaborate, and if the weight of this particular preference is large, and the effect of the mismatched preferences of the collaborating agents can be kept low, a meta-stable social network can be formed. This dynamics supports the observation that networks arise partly because agents choose to associate with others who are similar to themselves in some significant respect (Lazarsfeld and Merton 1944).

The emergence of a large-scale alignment of an environment along a certain preference common to a group of agents thus expresses the identity of that particular component at a large scale, while constraining the mismatched preference of the symbiotic components of each agent to his immediate locality.  Hence, an interaction minimum of a large magnitude exists with the semi-stable formation of a large scale environment alignment along that preference in a number of agents located separately but within a group identity.  Such highly weighted preferences can be said to have attained its own identity at the social or group level, which is reflected in the consistent environment minimum.

A specific situation of such coherent alignment in a very highly weighted preference, and the consequent dynamics in an extreme case, is described below.

\section{Altruism from Component Affiliation}
Experiments indicate that altruism need not be always based on expectation of a return (Bowles and Gintis 2004). However, models based on genetic kinship (Dawkins 1989) do not take into account all forms of altruistic behavior, e.g. those based on values.
We investigate special cases when a certain attribute/preference pattern component is in peril, and its survival or perpetuation instinct, from bonds created with matching patterns in other agents, may supersede the bond of the pattern with its own symbiotic components.

First, we look into the origin of a coherent faith based cluster so that the coherent expression of a faith among (almost) all members creates a faith identity expressed in a social level so that just as an agent has an individual identity as a component of the social cluster, the faith has a component identity within the cluster as expressed within agents coherently.

In our specific example two or more clusters with independent faith may exist in different localities, and when posed with competition (see Shafee 2008 for a discussion of competition based components among identities) over specific scarce common resources for contradictory preferences, the clusters may try to eliminate each other by realigning (converting) or eliminating agents who are members of the other clusters.

In such situations, when a cluster is in peril ({\it i.e. losing}), the rational choice of an agent is to defect, and save himself by converting to the other faith.

We present a specific scenario, when an individual agent is given the choice of saving himself (his entire locally coupled personal identity) and risking his life by not converting. Real world data show that in specific cases such irrational {\it altruistic} martyrdom exists (see ?????????????????????????)

In the scenario presented below, the agent chooses to sacrifice himself in an altruistic manner. The expressions of identities and the intersection of the faith component within each individual agent's identity and the social identity plays a vital role in this scenario. Identities in each case give rise to local interaction potential minima, and the intersection of identities come into play when one possible minimum is chosen over the other if only one of the two can exist within the agent's own frame.

In (Shafee 2008), an agent was given an identity because of a strongly coupled localized group of diverse variables.  The stability (meta-stability) of the identity derived from the local interaction minimum derived from was based on the symbiotic/match-based interactions among the components.  These highly localized components were semi-closed in the sense that they also coupled with neighboring environment states within interaction range, and states of other agents.

In (Shafee 2009) it was shown how a rational agent could be derived when the local internal couplings are strong and the external couplings provide random matches/mismatches.  However, the utility of an agent based on a preference, was also stated as modified after the zero-th rational order based on the commonality with other agents in extended group identities.

We first derive the concept of this larger identity from the point of view of couplings and stability, and then discuss a special case with coherent faith based couplings among agents within a common environment forming a strong extended identity.

The modified identity of the agent would include interaction terms in all three of his interaction Hamiltonian components, namely $H_{self}$, $H_{env}$ and $H_{agent-agent}$.  Hence interacting states or variables in the local environment and other interacting agents would also give rise to second and higher order identities.  However, because of the random match/mismatch among these states, and the fact that other agents would come in their own locally coupled bundle of states, many of them not optimal as interaction terms with the agent's own states, these {\it external} components of identity would be dynamic.  The agent himself modifies states within his own {\it self} array in order to adapt, so that a change is necessary to maintain the entity (local minimum) as a whole.  However, because of the fixed nature of many of his own states (the central dogma of molecular biology, so that information flows from genes to synthesized proteins, see Crick 1970), many of the internal states are fixed and coupled together in a massive manner, and disconnecting one component may threaten the integrity of the entire structure (see Shafee 2008).

We now consider a special case, when a group of isolated agents, connected within a local environment and experiencing very similar events, giving rise to similar sets of norms, may behave in a very different way.

A common set of experiences may give the agents a common set of norms at a time point (see e.g. Gordon 2003 for how Japan came together after World War II. If these people are genetically close, they also share a large subset of genetically fixed preferences in the form of locally preserved alleles (see Cavalli-Sforza and Edwards 1967 about genetic drifts and Nei 2005 for population bottlenecks).  Let us say that these agents share a common group of preferences or norms bundled as a faith that is brainwashed into a majority of them (see Shafee 2009b).  Now if these induced preferences or norms are connected with the agents long enough so that the local environment is largely aligned along the faith, the coupling of each agent with the similarly aligned local environment is also large.

Now we consider each agent's array split into three distinct parts based on his own genetic individuality and the effect of the faith.
1. preferences aligned along the faith :$P_F^i$, because of the connections of the faith with those preferences and the weight of the faith.  As time goes by, these preferences get coupled with the local environment because of experience and alignment of the environment along the faith. Agents in the network share a large overlap of these preferences.
2. individuality components: are the agent's own set of preferences that are not related with the faith.  These states may have random matches/mismatches with the local environment, network agents' states.
3. buffer variables: act as a buffer between the faith preferences and the individuality preferences.  They may come in two possible states, one optimal for the faith preferences and the other optimal for the individuality components so that triggering one (because of optimality in connection) blocks the other.
Just as within the agent, the three types of faith-related components are coupled to form the agent's own identity, the faith and $P_F^i$ components have a large degree of match-based coupling with other agents and also the local coherent environment, which is aligned along the shared faith component.
Hence, the coherence in alignment produces a network-spanning identity based on that faith.

The faith thus intersects the agent's individuality as it couples socially to form its own local interaction minimum based identity.

We now study the behavior of the buffer states, especially when the two intersecting identities come into play.

\subsection{High Priority Beliefs}
We start with a high priority pattern, e.g. a value that has high priority in an agent, so that an agent strives to connect to the environment with that preference with a high weight.  This component may be created by repeated historical interactions within a group of agents and the common environment, or this can also be introduced or imagined initially as a virtual term to stabilize some local critical components.

This belief, faith or norm component, $F$, may have competing components in other agents so that at the macro level, the change in the environment depends on the winning component or belief so that a certain number of agents must have this component in order for an environmental change to take place.  Or else, the belief loses and no alignment of the environment takes place, so that chaos prevails.

In situations like this, in order for the environment to be aligned in a certain way, the faith component must exist in a certain number of agents so that all these agents also have a group of preferences within them shifted on account of this belief, so that they all coherently invest in aligning the environment along that belief component's direction.  In such a situation, the belief component might have a value such that losing the other variables in order to preserve that belief component may be favorable, since the belief component, which has a high weight in the agent and his expressions, can be perpetuated at the cost of the other variables of the agent.

 Let $[F]$ be the group of belief norms originally shared by the agents.  Let [$P_i^F$] be the group of preferences or norms within the $i^{th}$ agent that are aligned along $[F]$.  Similarly, the [$E_i^F$] are the local environment states that are aligned along $[F]$.  Each agent's individuality arises from [$P_i^R$], which are the remaining preferences which are unrelated to $[F]$.  The buffer states as mentioned in the last section are given by [$B_i$], which can exist in the states [$B_i^F$] optimal to $[F]$ or in [$B_i^R$], which are optimal to the remainder states of the agent.  The agents' genetic states are given by [$G_i$], and if the cluster is reproductively isolated, [$G_i$] has overlapping optimal components along [$E_i$](adaptation within the locality because of mutation arising within the local environment and selected), expressed as [$G_i^E$], and also a large degree of overlap with [$G_j$] because of similar alleles in a small homogenous population. Let the degree of genetic overlap be $[G_o]$. If the agents are in a small community, $[E_i] \sim [E_j] \sim [E]$.

Each agent can thus be said to be coupled to the attribute $[F]$, which is also coupled to [E].  The couplings of the exact matches of $[F]$ among the $N$ agents in the cluster, and the couplings of these $N\space [F]$ terms and the respectively coupled [$P_i^F$] terms with the environment $[E]$, which is now an optimally coupled set of variables in the local environment, form an optimally interacting identity, and hence a social identity based on $[F]$.  Now each of these [F]s in an agent ($[F_i]$), are coupled to the rest of the agent's state because of the internal couplings within the agent locally.  However, the introduction of a homogenous [F] term among all agents (that help the agents maintain a favorable environment together) and the difference in individuality posed by the mismatched variables of $[P_i]$ induced along [F] within $[P_i^F]$ and the consequent remainder of individuality terms cause the internal local couplings within the agent to be slightly less optimal, in order to accommodate optimality of the environment by aligning with the other agents of the network along [F].  However, if there is a large degree of genetic overlap among agents and a large number of norms aligned in the same direction because of repeated similar social interactions and experiences, $[P_i^R]$ may become less prominent.

Now the variables belonging to $[B_i]$ attain their state by optimizing the local interaction Hamiltonian.  However, these variables face the dilemma of being coupled to both the social identity and the agent's remaining individual identity.

If a situation arises such that $[F]$ and $[G_i^R]$ and $[P_i^R]$ are competing in the sense that converting $[F]$ within the agent to a different state $[F']$ is needed to maintain the agent's couplings within $H_{self}$, the alignment of $[B_R]$ will be along the future optimal Hamiltonian local to it.  The couplings of the massive terms of $[P_i^F]$ and $[G_o]$ with the social identity, which, if massive, together with the coupling of $[F]$ with massive terms in a network may cause the agent's local interaction Hamiltonian to become less optimal than the present case if $[F]$ is flipped to $[F']$.  However, if attaching to $[F]$ can maintain a possible future local minimum (see Shafee 2008 for future decisions) for the social identity coupled with $[B_i]$, the alignment of $[B_i]$ may choose the social identity of [F] at the local identity of the agent, which may appear as irrational behavior. The loss of the variables in $[B_i]$ when the internal couplings of the $i^th$ agent are lost, are similar to the realignment of components of agent within the agent's identity in order to maintain the lowest aggregate local minimum.  This realignment at the cost of one agent comes within the identity of the social interaction potential, that is also local to the agent because of large degrees of coherence.

\section{Conclusion}
In this paper, we have discussed the role of the expression of components in an agent within the expression of the identity of an agent.  The concept of patterns and the preservation of patterns were discussed. Stability was discussed from the point of view of perpetuating pattern states, and identities evolving and perpetuating in different scales were discussed. A specific case was studied when a component identity played a game with an agent identity in terms of the preservation of the local potential minimum. The model thus extends the concepts introduced by the idea of interaction potentials in the social network to further explain some {\it puzzling} social phenomena.

\section*{Acknowledgement} The author would like to thank Prof Douglas White and Prof Bertrand Roehner for their continued encouragement.  She would also like to thank Prof M. Fisher for feedback on an earlier related paper.

\section*{Reference}

Alder EM, Cook A, Davidson D, West C, Bancroft J, Hormones, Mood and Sexuality in Lactating Women, British Journal of Psychiatry (1986), 148, 74-79

Anderson JA , An Introduction to Neural Networks, MIT Press (1995).

Axelsson J, Åkerstedt T, Kecklund G, Lindqvist A, and Attefors R,Hormonal changes in satisfied and dissatisfied shift workers across a shift cycle J Appl Physiol 95: 2099-2105, 2003.

Black F,  Scholes M, The Pricing of Options and Corporate Liabilities, J. Pol. Econ. 1973; 8 .1 (3): 637-654

Brett D. Pospisil H, Valcárcel J; Reich J; Bork P Alternative splicing and genome complexity. Nature Genetics (2001) 30: 29–30.

Cavalli-Sforza LL and Edwards AWF "Phylogenetic analysis: Models and estimation procedures". Evol. (1967) 21 (3): 550–570

Crick, F. (1970): Central Dogma of Molecular Biology. Nature 227, 561-563

Edwards S and Anderson P, Theory of spin glasses, J. Phys. F. 1975; 5: 965-974.

Finn, R. (2000). Organ Transplants: Making the Most of Your Gift of Life. Sebastopol: O'Reilly \& Associates.

Gordon A A Modern History of Japan. New York: Oxford University Press, 2003

Griffiths AJF, Miller JH, Suzuki DT, Lewontin RC, Gelbart WM (1993). "Chapter 5". An Introduction to Genetic Analysis (5th ed. ed.). New York: W.H. Freeman and Company. ISBN 0-7167-2285-2

Hopfield JJ, Herz AVM Rapid local synchronization of action potentials: toward computation with coupled integrate-and-fire neurons, Proc. Natl. Acad. Sci. {\bf 92}, 6655 (1995)

 Helbing D. and  Schreckenberg M. (1999) Cellular automata simulating experimental properties of traffic flows. Physical Review E 59, R2505-R2508.

 Lazarsfeld PF and Merton RK, Friendship as a social process: A substantive and methodological Analysis, in  Freedom and Control in Modern Society. Ed. M. Berger et al. Van No strand; 1954. p.18.

 Lebedev, M.A., Carmena, J.M., O'Doherty, J.E., Zacksenhouse, M., Henriquez, C.S., Principe, J.C., Nicolelis, M.A.L. (2005), Cortical ensemble adaptation to represent actuators controlled by a brain-machine interface. J. Neurosci. 25: 4681-4693

 Minzenberg MJ, , Antonia RG.  S New, Vivian Mitropoulou, Rachel Yehuda, Marianne Goodman, Diedre A Reynolds3, Jeremy M Silverman1,Blunted Hormone Responses to Ipsapirone are Associated with Trait Impulsivity in Personality Disorder Patients, Neuropsychopharmacology (2006) 31, 197–203

Nei M. "Bottlenecks, Genetic Polymorphism and Speciation". Genetics (2005) 170 (1): 1–4.

Newman M E. J,   Barabási, AL Watts DJ. The Structure and Dynamics of Networks, Princeton University Press, 2006

Nicolopoulou-Stamati  Congenital Diseases and the Environment Series: Environmental Science and Technology Library P, Hens, L.; Howard, C.V. (Eds.) (2007) Vol. 23

Peterson JB and Carson S, Latent inhibition and openness to experience in a high-achieving student population. Personality and Individual Differences. 2000; 28: 323–332

Reichardt, J and White D, Role Models for Complex Networks Proceedings of the National Academy of Sciences of the United States of America; 2007, submitted.

Sherrington D   and Kirkpatrick SK, Solvable Model of Spin Glass,  Phys. Rev. Lett. 1975; 35: 1792-1796
Black F,  Scholes M, The Pricing of Options and Corporate Liabilities, J. Pol. Econ. 1973; 8 .1 (3): 637-654

Shafee F. Network of Perceptions, submitted  (2009b)

Stamma S,Ben-Ari S, Rafalska I,  Tanga Y, Zhanga Z,Toiber D, Thanaraj TA,  Soreqb H, Methods of Alternative Splicing, Gene 344, (2005) 1-20

Toga  AW and Thompson PM  Genetics of Brain Structure and Intelligence, Annual Review of Neuroscience Vol. 28: 1-23 (2005)

Walton M, Mice, men share 99 percent of genes, CNN, Wednesday, December 4, 2002

\end{document}